\documentclass[journal]{IEEEtran}
\usepackage{graphicx, soul}
\usepackage{soul}

\usepackage[colorlinks]{hyperref}

\usepackage{amsmath}
\usepackage{xcolor}
\usepackage{footnote}

\begin{document}
%
\title{A Speculative Study on 6G}
%
%
%

\author{Faisal Tariq,~\IEEEmembership{Member,~IEEE,}
        Muhammad R. A. Khandaker,~\IEEEmembership{Senior Member,~IEEE,}
        Kai-Kit Wong,~\IEEEmembership{Fellow,~IEEE,}
        Muhammad Imran,~\IEEEmembership{Senior Member,~IEEE,}
        Mehdi Bennis,~\IEEEmembership{Senior Member,~IEEE,}\\
        and~M\'erouane Debbah,~\IEEEmembership{Fellow,~IEEE}

}

%
%

\markboth{IEEE XX Magazine,~Vol.~XX, No.~X,}%
{6G Wireless System}
%



\maketitle

\textcolor{blue}{\emph{[Note: This  work  has  been  submitted  to  the  IEEE  for  possible  publication. Copyright  may  be  transferred  without  notice,  after  which  this  version  may no longer be accessible]}}

\begin{abstract}
While 5G is being tested worldwide and anticipated to be rolled out gradually in 2019, researchers around the world are beginning to turn their attention to what 6G might be in 10+ years time, and there are already initiatives in various countries focusing on the research of possible 6G technologies. This article aims to extend the vision of 5G to more ambitious scenarios in a more distant future and speculates on the visionary technologies that could provide the step changes needed for enabling 6G.
\end{abstract}

\begin{IEEEkeywords}
B5G, Emerging technologies, 6G.
\end{IEEEkeywords}

%
\IEEEpeerreviewmaketitle

\section{Introduction}
\begin{figure}[ht!]
\centering
\includegraphics[width=0.8\linewidth]{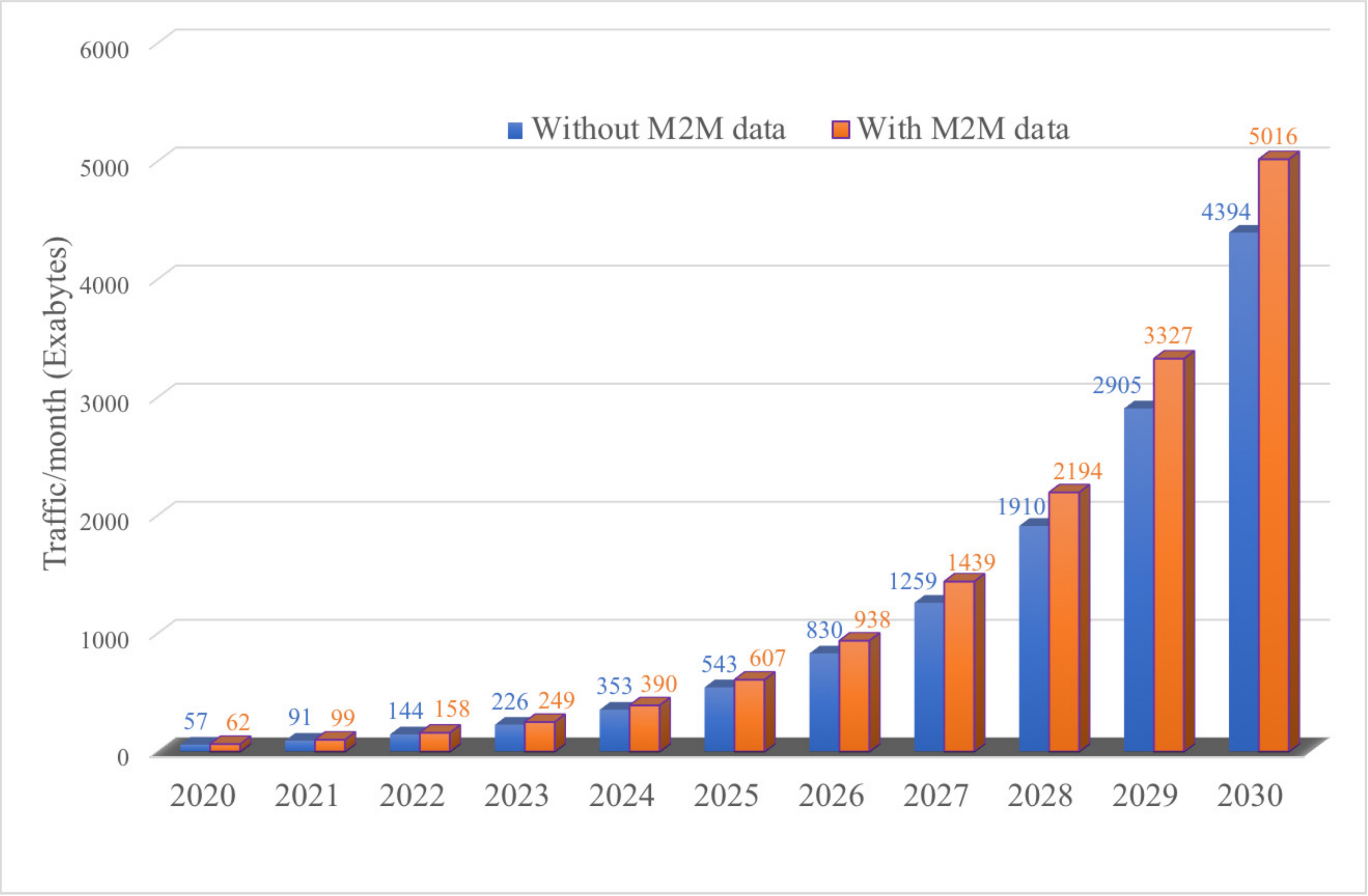}
\caption{Global mobile data traffic forecast by ITU. Overall mobile data traffic is estimated to grow at an annual rate of around $55\%$ in 2020--2030 to reach $607$ exabytes (EB) in 2025 and $5,016$ EB in 2030. (Source: Cisco)}\label{fig:traffic}
\end{figure}

The last decade has witnessed a never-ending growth in the global mobile data traffic, expecting to experience a 23 times increase in 2021 in data volume compared to the entire global Internet traffic in 2005. The International Telecommunication Union (ITU) predicted that the trend of exponential growth will continue and by 2030, the overall mobile data traffic will reach astonishingly 5 zettabytes (ZB) per month, as illustrated in Fig.~\ref{fig:traffic}. The fifth generation (5G) is the latest attempt that brings mobile communications technology up to speed to meet the requirements for the next 10 years. It is expected that 5G will reach its limits by 2030 and the chase continues.

Coupled with the rises of Internet-of-Things (IoT), massive machine-type communications (MTC) and etc, 5G is so much more than traditional cellular networks, and thrives to, among a few other key performance indicators (KPIs), achieve $1000\times$ increase in capacity compared to the 4G networks, delivering gigabits per second compared to megabits per second in 4G \cite{JAndrews14}. European Telecommunications Standards Institute (ETSI) published a document on 5G scenarios and requirement for access technologies where the target peak rate is reported to be 10Gbps in the uplink and 20Gbps for the downlink (3GPP TR 38.913). Further details for the various KPIs can be found in Table \ref{5g6g} where we also speculate on how the requirements may become in the possible 6G networks against 5G.

\begin{table}[ht!]
    \centering
\begin{tabular}{ |c|c|c|} 
 \hline
{\bf Characteristics} & {\bf 5G} & {\bf 6G} \\ 
\hline\hline
Individual data rate & $1$ Gbps & $100$ Gbps \\ 
\hline
DL data rate & $20$ Gbps & $>1$ Tbps \\ 
\hline
U-plane latency & $0.5$ ms\footnotemark{} & $<0.1$ ms\\
\hline
C-plane latency & $10$ ms & $<1$ ms \\ 
\hline
Mobility & up to $500$ km/h  & up to $1000$ km/hr\\ 
 \hline

DL spectral efficiency & $30$ bps/Hz & $100$ bps/Hz\\ 
 \hline
Operating frequency& $3-300$ GHz\footnotemark{} & up to $1$ THz\\
\hline
\end{tabular}
\caption{KPIs for 5G versus 6G.}\label{5g6g}
\end{table}


The myriads of emerging massively data-intensive use cases and applications such as multi-way virtual meeting, virtual and augmented reality (VAR) based gaming and remote surgery, holographic projection, etc., to name just a few, indicates that 5G will be unable to support these demanding applications ubiquitously and 6G will need to be called upon, similar to the case when multimedia services were introduced in 3G but full support is not achieved until 4G. Presumably, 6G will continue to benefit from many 5G technologies, but new technologies will certainly be needed to make the next step change.

The objective of this article is to provide an expert view on the most trending and novel research directions that have the potential to shape the 6G mobile technologies. Although the development of 6G is in an early stage and it is expected that some ideas will only emerge in later years, this visionary article takes a bravery approach to speculate on the possible enabling technologies and the revolutionary elements in 6G, and describe their features beyond the capability of 5G. 

The rest of this article is organized as follows. In Section~\ref{sec_vision}, we present our 6G vision and speculate on the requirements. Section~\ref{sec_ucases} then extends some 5G use cases to more ambitious scenarios that are expected to appear in 6G. We will discuss some key challenges in Section~\ref{sec_chall}, and present a few visionary technologies or research directions that may form key parts of 6G. Finally, we conclude this article in Section~\ref{sec_con}. 

\section{Vision}\label{sec_vision}
As 5G is entering the deployment phase, discussion for 6G is gradually taking momentum. 
It is still an early stage to formally define 6G, and any such discussion is more or less a speculation. Nevertheless, there is no doubt that 6G is taking shape. 
In this section, we attempt to be visionary and engrave our 6G vision, by highlighting several promising directions.

Building upon the 5G vision, 6G will continue to empower our cities to be super smart and fully connected with a plethora of autonomous services for mobile phones/tablets, IoT devices, driverless cars and many more. Large cities will see further penetration of flying taxis which is already operating in limited scale in various cities like Dubai. The command and control as well as connectivity requirement of those flying taxis and cars will be unprecedented. One reality of 6G is that 6G will be empowered by artificial intelligence (AI) in almost all levels, from network orchestration and management to coding and signal processing in the physical layer, manipulation of smart structures, and to data mining at the network and device level for service-based context-aware communications, etc.

\begin{figure}[ht!]
\centering
\includegraphics[width=.8\linewidth]{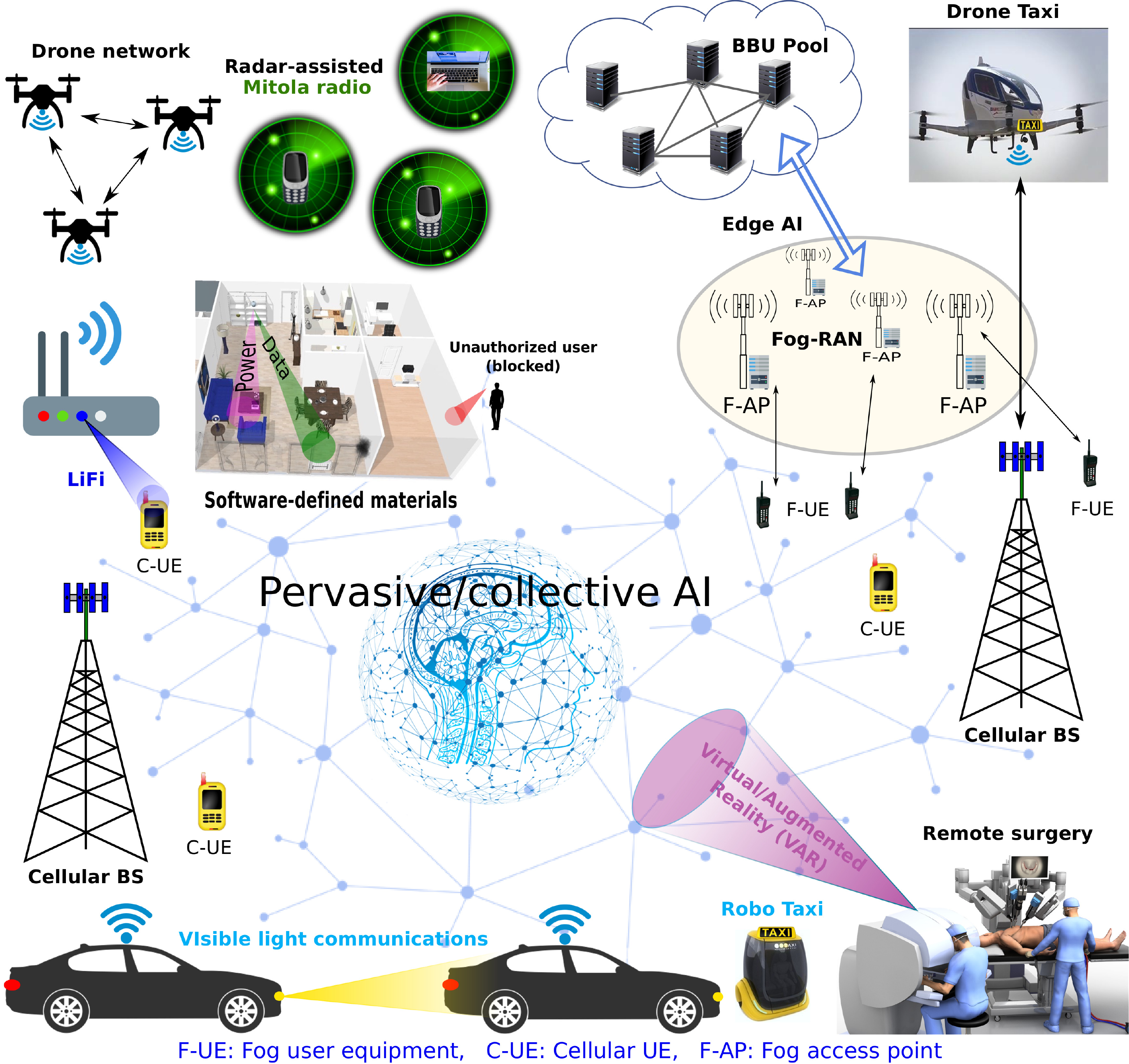}
\caption{The vision of 6G.}\label{fig:6g_vision}
\end{figure}

Everything around us will be very intelligent, giving rise to the concept of Internet of everything (IoE), with an enormous amount of data and information. Few will disagree that AI will be an integral part of 6G due to the availability of massive actionable data and advances in computation capability. 
Recent interest has also shifted to edge caching and fog radio access networking (fog-RAN), which brings contents closer to user equipments (UEs), allowing for much lower latency and power consumption. In 6G, we anticipate to see AI in operation with distributed training at the network edges including small-cell base stations (SBSs) and UEs, which is still an open problem. In contrast to the use of conventional AI algorithms on mobile network data in 5G, 6G will realize the notion of {\em collective AI}, a step-up from the current AI techniques that will address the coexistence of multiple distributed mobile radio learning agents for individual as well as {\em global} benefits.

At the device level, another reality is that radar technologies will be integrated with mobile communication technologies to provide all-round contextual information from the small scale to medium scale to assist communications. In 6G, it is likely to see that physical-layer security finally thrives to provide a layer of defence, in addition to cryptographic techniques, for a variety of devices and machines with different capabilities. Contextual information obtained by radars installed at mobile phones, tablets or any IoT devices will make possible physical layer security approaches to be adopted, not to mention many other uses of contextual information in the application level. Devices will also be much smarter, empowered by AI trained by the behavioural data of the environment from radars.

On the other hand, after 30 years of effort, 6G will finally see Mitola radio, a.k.a.~cognitive radio, reach its full potential. Despite the enormous interest after Mitola {\em et al.}~first introduced the concept back in 1999 \cite{mitola-99}, little has been achieved so far although there are some basic shared spectrum technologies for 5G new radio (NR).
There is still a huge gap from being the cognitive radio Mitola envisioned. The 6G Mitola radio will see self-regulating societies of mobile radios for fair as well as efficient coexistence and facilitate seamless mobile convergence across LTE, Wi-Fi and other networks.  

Apart from these, 6G will utilize smart structures to provide an additional degree of freedom (DoF) to improve the wireless links, delivering an unprecedented capacity. In the large scale, smart reflective surfaces will be installed in buildings \cite{Hum-14}. The smart surfaces will effectively increase the antenna aperture to collect as much radio signals as possible that had not been possible before for improved energy and spectral efficiency. In the smaller scale, 6G will also see flexible antenna structure possible at UEs. Early results on fluid antennas in \cite{Tong-17} revealed a whole new possibility for designing wireless communications systems. Also, metamaterial-based antennas may also be implemented to make even more compact wideband antennas. 
Such intelligent structures seek to engineer the environment to cater a variety of applications, for example, to improve link quality, block interference, enhance privacy and security, avoid adversarial attacks and many more.

There is also glimpse of successes in other emerging areas which are not yet making much of an impact in 5G but could become reality in 6G, including wireless power transfer (WPT) and RF energy harvesting, optical wireless communications or Li-Fi. Furthermore, there is a possibility that 6G will be more than wireless, and need to handle coexistence of traditional mobile communications and interconnects inside PCs \cite{Mak-16}, as many-core PCs may use digital surface-wave communications for interconnects which might occupy the same band as 6G. Overall, we foresee that AI will penetrate all levels and be the signature for 6G for smarter and more powerful networks.

In terms of the requirements in 6G, the consensus seems to be that the data rate will race to 1Tbps to enable autonomous management of various activities in the future smart city. For individual users, data rate is expected to increase from 1Gbps in 5G to at least 10Gbps per user, and up to 100Gbps in some use cases, in the emerging 6G systems. 6G is also expected to integrate with satellites for providing global mobile coverage. Volume spectral efficiency (in bps/Hz/m$^3$), as opposed to the often used area spectral efficiency (bps/Hz/m$^2$), will be more suitable in 6G to properly measure the system capacity in a three dimensional operating space.
Ultra-reliable low-latency communication (URLLC), one key feature in 5G NR, will again be a key driver in 6G pushing the limit further to require latency of less than 1ms. Energy efficiency will be extremely important to prolong the battery life of UEs. The KPIs for 6G in comparison with 5G are shown in Table \ref{5g6g}.

\section{Use Cases}\label{sec_ucases}
Most of the use cases in 6G will evolve from the emerging 5G system based applications in terms of functionalities and quality of experience. When the enablers of 6G systems gradually become available, the applications will follow through by adding further performance enhancements and new use cases. In this section, we use three popular 5G use cases as evolutionary examples to describe what 6G may bring beyond the capability of 5G. A comparison of a list of 5G and 6G use cases is provided in Table~\ref{tab_ucases} for quick reference.

\begin{table}[ht!]
    \centering
\begin{tabular}{ |c|c|c|} 
 \hline
 
{\bf Use case} & {\bf 5G} & {\bf 6G} \\ 
\hline\hline

Centre of gravity & user-centric & service-centric\\ 
\hline

Ultra-sensitive applications & not feasible & feasible\\ 
\hline

True AI & absent & present\\
\hline

Reliability & not extreme & extreme\\
\hline

VAR & partial & massive scale\\ 
 \hline

Time buffer & not real-time & real-time\\ 
 \hline
 
Capacity & 1-D (bps/Hz) or & 3-D (bps/Hz/m$^3$)\\
& 2-D (bps/Hz/m$^2$) & \\
\hline

VLC & no & yes\\ 
 \hline

Satellite integration & no & yes\\ 
 \hline

WPT & no & yes\\ 
 \hline

Smart city components & separate & integrated\\ 
 \hline

Autonomous V2X & partially & fully\\ 
 \hline
 
\end{tabular}
\caption{Comparison between 5G and 6G use cases.}\label{tab_ucases}
\end{table}

\vspace{-.3in}
\subsection{Haptic Communication for VAR}
Haptic communication adds the sense of `touch' to traditional audio visual communication over the Internet, and is the key to unlock the potential of VAR \cite{MBennis1}. This will have massive impact on different economic sectors including manufacturing, education, healthcare, and smart utilities. Various appealing services have been discussed in the development of 5G, such as remote surgery, holographic communication, networked games in simulated realities, and etc. \cite{VanDen}. These services will require varied degree of latency and reliability. 

Although URLLC in 5G can deliver latency-critical applications (e.g., 10ms), some ultra-sensitive applications such as remote surgery will require latency to be less than 1ms which is not yet achievable in the upcoming 5G systems. Another most-talked application in 5G is holographic communication, which however may come in limited form in 5G and is only possible with dedicated network resources and under limited or no mobility conditions. 5G is unlikely able to cope with a massive number of holographic communications with reliable performance. Other high-demanding VAR applications such as virtual meeting room and holographic projection also require massive amount of real-time data transfer over the air, and will need 6G to meet the end-to-end latency requirements. 


\vspace{-.1in}
\subsection{Massive IoT Integrated Smart City}
The jargon `smart city' describes the concept where a city greatly improves the quality of life (QoL) for the people in it by optimizing its operations and functions using the available infrastructures to sense, detect, analyze and act by integrating the core components that run the city. IoT is the enabler for realizing smart cities, and it is one of the main 5G objectives to make smart city a reality. However, in 5G, a city can only be fragmentally  smart, meaning that the major components such as utilities (i.e., electricity, water and waste, etc.), healthcare and monitoring, and transportation networks are individually smart but considered separately.  Compared to 5G, 6G will take a holistic approach in an integrated fashion for a truly smart city. We elaborate on three major smart city applications. 

\subsubsection{Smart Homes}
Over the last decade, the efforts in smart utilities focused largely on smart energy grid due to ease of integration and particularly the rapid progress in smart meter deployment. However, the progress for other utilities has been much slower. Despite this, IoT technology will soon mature and liberate all the elements in homes, and make them smart. The challenge of course will be to support the massive data rate at homes and provide security protection for personal data. 6G will be tasked to provide the necessary infrastructure for all the data intensive services. Also, 6G is expected to integrate fully with AI for autonomous decision making in homes.

\subsubsection{Connected Vehicles and Autonomous Driving}
The automotive and transportation industries are experiencing a generation change, partly due to the connectivity and networking capability offered by 5G and beyond systems. Several standards have recently been developed such as dedicated short-range communications (DSRC) and vehicle-to-everything (V2X) to enable smart vehicular systems. With AI and extreme data rate, autonomous and connected vehicular technologies will reach a new height. Note that massive amount of data will need to be shared amongst vehicles to update live traffic and real-time hazard information on the roads, and to provide high-definition 3D maps. All those information will be processed by AI in the autonomous command and control transportation network. As vehicles generally move at very high speeds, the network needs extremely low round-trip time for communication. It is highly likely that  V2X technologies will not mature by the 5G cycle and its full potential will only be realized in 6G.

\subsubsection{Smart Healthcare}
Ageing population is putting a huge burden on the healthcare system. A number of IoT based solutions for ubiquitous health monitoring have been developed to monitor various health indicators such as temperature, heart-rate, glucose level, blood pressure and automatically report those medical data back to the relevant department. As this is a very sensitive and delicate issue, a high level of security, reliability, and ubiquitous availability of communication system as well as network infrastructure must be ensured. In addition, smart healthcare will be aimed at giving the same experience to patients as if they are diagnosed by doctors in person, which means that secure high-definition video conferencing will be needed over the air. 6G will be needed for smart healthcare to truly take off and gain popularity for wide acceptance.


\vspace{-.1in}
\subsection{Automation and Manufacturing}
Massive incorporation of robots into automation and warehouse transportation is vital for industry growth. The emerging concept of Industry X.0 aims to enhance the Industry 4.0 by exploiting social, mobile, analytics and cloud (SMAC). The radio environment with a very complex network comprised of hundreds or thousands of robots is a challenge. 6G will fully support the Industry X.0 revolution by offering massive URLLC as well as massive IoT and embedded AI capability.  


\section{Challenges}\label{sec_chall}

As seen in Table \ref{5g6g}, 6G looks for several orders of magnitude improvements over 5G in all aspects. Here, we highlight a few hurdles, where major efforts will be needed. Some of them are already being looked at and partially addressed in 5G.

%
%
%

\subsection{Access Network for Backhaul Traffic}
The recently formed ITU focus group technologies for networks  2030 (FG NET-2030) raised concern that fixed access networks capabilities are already lagging behind emerging 5G systems. It is anticipated that  access networks for backhahul traffic will struggle to cope with unprecedented data growth and other quality requirement unless necessary steps to boost the research effort are initiated. There are plenty of spectrum available in higher bands, and one such option is the D-band where 60GHz spectrum is available. 
Free space optical communications and quantum communications are the hopefuls for 6G backhaul to meet the requirements. However, they are still in an early stage of development and how these technologies, even if they exist, will be integrated with the other types of network equipment also needs to be addressed.

\subsection{Sub-Millimeter Wave and THz Frequencies} 
An obvious way of supporting massive increase in the data rate requirement is to increase the bandwidth. Initial discussion indicates that frequencies in the range of THz and above will be considered for 6G as in those bands there are plenty of free bands suitable to satisfy this requirement. A recent study indicates that 5G spectrum may not exceed 140GHz due to a number of challenges including the lack of understanding of channel and propagation modelling, device inability to operate at such high frequencies, and etc. 
In contrast, 6G will utilize spectrum beyond 140GHz with particular application in very short range communication or `whisper radio' \cite{Rappaport1}. 
However, the susceptibility of the THz band to blockage, molecular absorption, sampling and circuits for A/D \& D/A conversion and communication range is among the major challenges that need to be addressed in the coming years. Another issue is that at higher frequencies, the antenna size and associated circuitry become miniaturized and are difficult to fabricate on chip while ensuring noise and inter-component interference suppression. On the other hand, the exact propagation characteristics in these bands is not well understood, although a few recent attempts to address these bottlenecks have reported encouraging results. In one successful example, it shows that it is possible to design a CMOS based modulation circuit operating at 300GHz to obtain signicant performance gain.


\subsection{Fog Networking and Mobile Edge Computing}
Fog networking and edge computing have been introduced in 5G to greatly shorten the UEs' distance with the serving base stations and service/application content servers, respectively. However, as mentioned earlier, edge caching breaks the network into a distributed cloud structure where training data reside at the network edges, which handicaps AI techniques to be fully functional. It is inevitable that the network is going to move towards even smaller cells for more capacity and less latency in 6G, and this situation will exacerbate.

\subsection{Resource as a Service (RaaS)}
The emergence of software defined networking (SDN) and network function virtualization (NFV) facilitates a move towards service oriented and integrated resource distribution which is known as RaaS. This has given rise to the concept of network slicing to create virtual networks over the physical infrastructure. It allows mobile operators or service providers to allocate virtual network resources to meet specific service needs. In 6G, programmable metasurfaces and software-defined materials will likely be parts of the network resources. Thus, one trend of development for NFV in the 6G cycle will include network slicing with software-defined materials and programmable metasurfaces, from machine learning-enabled cloud random access network (C-RAN) to fog-RAN. 


\subsection{Dynamic Topology}
In light of network densification, each end node will have multiple options to connect and user association decision will have a great impact on the interference pattern. This coupling with dynamic dronecells formed by unmanned aerial vehicles (UAVs) and fast moving vehicular networks, will change the interference dynamics rapidly. Modelling this new interference dynamics and ensuring that the network takes full advantage of the flexibility and adaptivity of cell shapes as well as topology will be a priority of 6G. New mathematical tools will need to be discovered to allow such analysis and optimization.


\subsection{Device Capability}

Every generation of mobile communications has been defined by the UE capability, and this will be more so for 6G because 6G will be AI-led and require high computational power to run the AI algorithms. UE will be more power hungry than ever. Energy efficiency at the device level will once again be a KPI in 6G. Furthermore, for devices to operate like a Mitola radio, new materials and design concepts will need to be sought to break the physical limits of UE and yet operate at wider frequencies with great diversity and multiplexing gains as well as intelligence.  Conventional transceiver components are mostly based on semiconductor materials like Silicon (Si) and Gallium Arsenide (GaAs). These devices are not energy efficient and produce excessive heat. Since the maximum operating frequency of CMOS transistors has not improved from around 300GHz in the manufacturing process after $65$ nm in line with miniaturization, 300GHz-band amplifiers with a CMOS integrated circuit are extremely difficult to realize. This means that they are not capable of supporting computationally intensive and ultra fast applications for 6G. Thus, new devices need to be designed based on new materials which have the characteristics to support the need for emerging systems.



\subsection{Blockchain Enabled Security and Authentication} 
With over 50 billion UEs and IoT devices connected everywhere with different levels of capability, 6G will need a holistic approach to secure the sheer volume of mobile data across a diverse set of platforms and comply with the strict privacy and security requirements. Blockchain technology will likely play a major role in securing and authenticating future communication systems thanks to the inherent advantage of the distributed ledger technology. Some promising use cases will include distributed security management for IoT, offloading in mobile edge computing, NFV and content caching.


\section{Key Enabling Technologies}\label{sec_enablers}
6G will look for another major leap from 5G, which would only be possible if breakthroughs could be achieved. Here we share some emerging ideas that could contribute significantly to enable 6G in the next 10 years or so. We intend to be more visionary than conservative so there is a likelihood that some ideas to be discussed in this section may not be fully ready in the 6G cycle but could appear as an enhancement of 6G.

\vspace{-.1in}
\subsection{Pervasive AI}
The most certain enabling technology for 6G has to be AI. Due to the advances in AI techniques especially deep learning and the availability of massive training data, recent years have seen an overwhelming interest in using AI for the design and optimization of wireless networks, and it is a consensus that AI will be at the heart of 6G (cf.~Fig.~\ref{fig:6g_vision}). In fact, recent success has motivated AI to form part of 5G although in 5G AI is only expected to operate in isolated areas in which massive training data and powerful computing facility are available. So far researchers have shown numerous successful examples of using AI on wireless communications, from physical layer designs such as channel estimation and precoding, to network resource allocation such as traffic control and cache storage management, to security and authentication, and to dynamic cell and topology formation and management, to fault prediction and detection, and etc. The list continues. It is reasonable to believe that some form of AI will be realized as part of 5G and AI will become more of the core components in 6G.



Deep learning, the most powerful AI techniques at present, is however based on deep neural network (DNN), which relies on training in a centralized fashion. Yet 6G is moving towards a more distributed architecture like fog-RAN handling millions and billions of end-to-end communications  anywhere around the world. The distributed cloud structure necessitates training to be done at the network edges and handicaps the operation of deep learning. Although the recently developed federated learning partly addresses this problem by allowing training to take place at distributed locations, this is more a distributed implementation for centralized learning, and communications between the distributed clouds and a central network manager is required \cite{Bennis-18}. Also, federated learning for optimization is much less powerful because updates in the user level are averaged before sending back to the central manager.

Nonetheless, the 6G Mitola radio is aiming at realizing true AI in an integrated manner from the device level all the way up to the entire network. The fine resolution of the specific needs and constraints of the UEs is as essential. The required optimization draws some similarity to players competing in a game, so game theory is expected to be useful although game theory by itself is not readily an optimization method. For 6G to succeed, an integration of AI and game theory will enable a truly distributed learning mechanism where multiple AI agents can teach and learn from each other by interactions. Collective AI is a related concept that has emerged recently to deal with the situation where multiple AI agents aim to achieve the same goal based on local training at each AI agent, with limited or no direct communication amongst the agents \cite{Jun-17}. We estimate that collective AI that can combine AI with game theory in an effective manner will bring the true brain power to 6G.




\vspace{-.1in}
\subsection{Radar-Enabled Contextual Communications}
Intelligence only shows if there is sufficient information to analyze. 
Radar technologies enrich environmental awareness for mobile UEs and IoT devices and enable context-aware communications to the level that has not been able to achieve before. This will give 6G radios the environmental awareness to empower AI at the device level. Combining the observations from radars with AI, UEs will be able to identify and localize potential eavesdroppers or adversaries by observations from radar and adapt their communications for enhanced protection using physical layer security approaches. The 6G Mitola radios will also store behavioural data of the environment, and can predict suspicious activities of malicious nodes. In addition, methods like physical layer authentication which relies on users' behavioural data will be possible. The general contextual data gathered by UEs will also assist the network to serve better by predicting UEs' next moves.

\vspace{-.1in}
\subsection{Cell-Free Networks} 
One hot topic at the late stage development of 5G is UAV wireless network which proposes to adopt flying base stations for providing mobile coverage in situations where there are no infrastructures or the infrastructures are heavily compromised due to disastrous events, and reconnaissance activities. In 6G, the full potential of UAV wireless networks or dronecells will be realized, and their application will be widely extended to mobilize the network resources to achieve cell-free networks where arbitrarily small latency may be obtained. To take full advantage of the fluid cells formed by UAVs, the optimization for resource allocation (including radio, energy and computing resources), trajectory, content caching and user association will be achieved jointly. Also, in 6G, UAVs will not only serve as flying base stations to provide radio coverage but also content providers and computing servers. There will be lots of synergy with other emerging technologies. For example, AI will take the network usage data to learn and dynamically find the best paths for the UAVs and optimize their other parameters. This will inevitably lead to  dynamic reconfigurations of the network topology. In addition, UAVs will benefit greatly from WPT technologies that can keep them moving all the time, while UAVs will also help support service-based network slicing.

\vspace{-.1in}
\subsection{Metamaterials based Programmable Radio Environment}
Despite the enormous success over the past decades, traditional antenna design techniques have seemingly reached their limits, and any further efforts are expected to result only mild incremental gains. Nonetheless, metamaterials-based antennas have been researched for nearly two decades, not yet making an impact in previous generations of mobile communications. This will change in 6G where metamaterials-based antennas will become the norm for UEs, permitting the massive MIMO technology to be employed at mobile phones as well. The maturity of metamaterials-based antennas will also make small-sized highly efficient wideband antennas possible, which gives the hardware flexibility that the 6G Mitola radio needs.

Another new form of antenna technology, which will come to light in 6G, is fluid antenna \cite{Tong-17}, made of conductive fluid, metal fluid or ionized liquid that can be shaped to any desirable form to suit the propagation environment. The fluidic structure breaks the boundary between the pre-defined antenna hardware and signal processing, and allows to optimize its position and shape for extraordinary diversity and multiplexing gains, while having the ability to reduce the electromagnetic fields (EMF) exposure by adapting to human gesture in the case of mobile phones, based on the environment and the needs at any given time. Presumably, a single software-defined fluid antenna can provide the rich diversity that only massive MIMO antennas could achieve, while enjoying the flexibility to alter its shape, size and position to fully utilize the surface of a UE. 

\begin{figure}[ht!]
\centering
\includegraphics[width=0.85\linewidth]{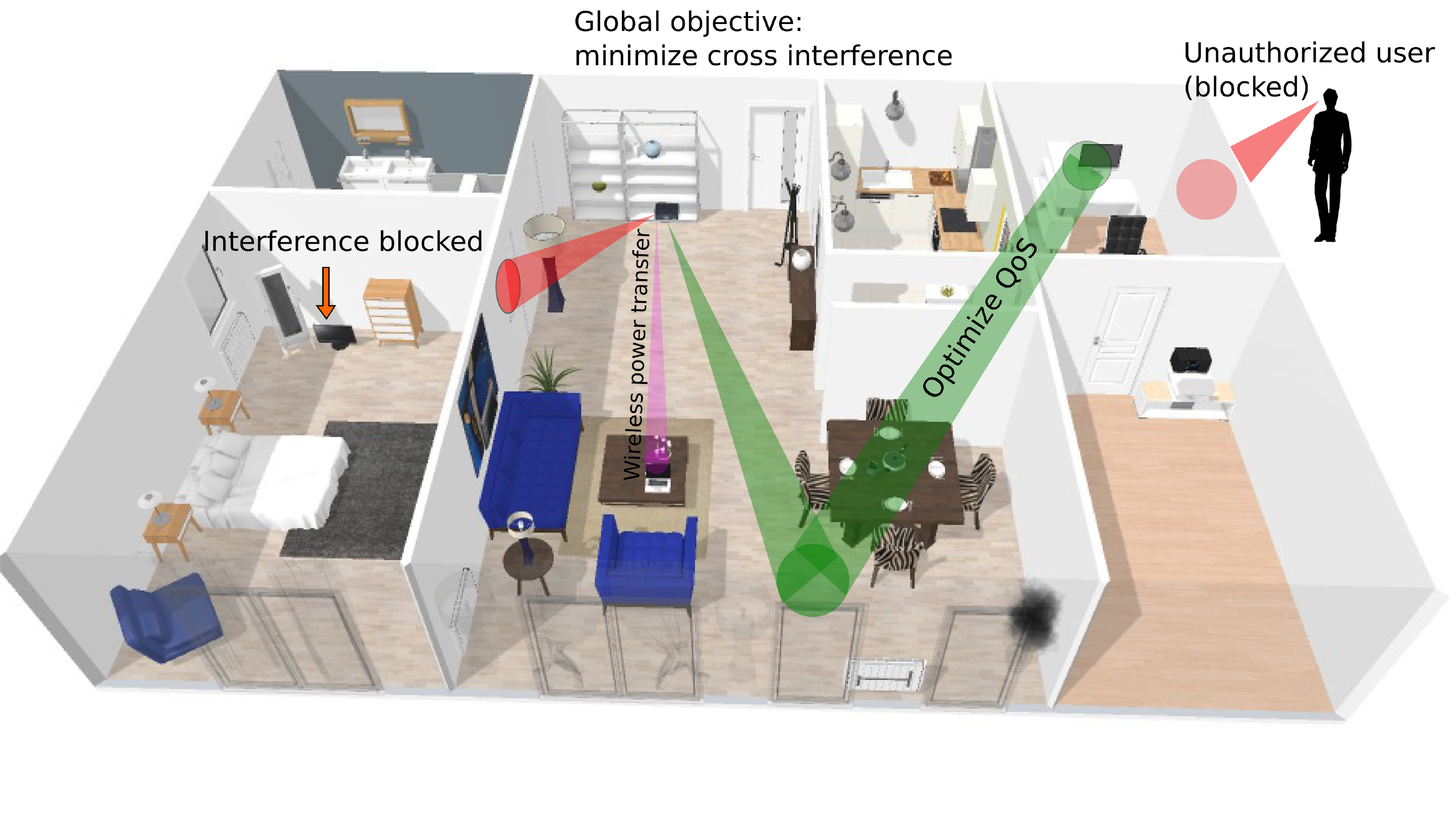}
\caption{An intelligent radio environment using SDM.}\label{fig:SDM}
\end{figure}

Software-defined materials (SDM) can actually be 
used to design large intelligent surfaces (LIS) to enable programmable wireless environment as well as for enhancing the coverage area of ultra-small cells 
\cite{CLiaskos}. By doing so, it is possible to control the propagation environment by altering its electromagnetic properties. For example, SDM can be laid on walls to provide insulation for unintended radio signals as shown in Fig.~\ref{fig:SDM} \cite{CLiaskos}. LISs on buildings or in indoor environments are predicted by many to make their mark in 6G.

Programmable metasurfaces indeed can do much more than simply altering propagation environments. They could replace the design architecture of wireless transceivers entirely. Recent results reveal that a metasurface can be programmed to vary the phase, amplitude, frequency and even orbital angular momentum (OAM) of an electromagnetic (EM) wave, effectively performing the modulation of a radio signal without a mixer, and RF chain \cite{Jin-19}. This technology will be quite disruptive and depending on its progress, we may see its operation as an enhancement of 6G, if clear benefits can be materialized.


\vspace{-.1in}
\subsection{Visible Light Communication (VLC)}
VLC is a special form of optical wireless communications and uses white-LED to encode data in the optical frequencies and some works have suggested that each link can achieve up to 0.5Gbps data rate \cite{haasICCW}, hence a good candidate to meet the data rate requirement of 6G. We speculate that VLC will be especially useful in vehicle-to-vehicle (V2V) communications where car's head and tail lights can be used as antennas for `control and coordination' data communication. VLC will also be useful in scenarios in which traditional RF communication is less effective such as in-cabin internet service in airplanes, underwater communication, healthcare zones and etc.

\vspace{-.1in}
\subsection{WPT and Energy Harvesting}
WPT does not play a key role in 5G but in 6G, it will finally shine for a number of reasons. First, communication distance will be much shorter, making WPT meaningful, because wireless networks continue to be denser plus the use of UAVs as base stations further reducing the distances. Also, UEs or any IoT devices in 6G will be more power hungry because of the huge computation demands for AI processing. On the other hand, energy scavenging from ambient RF signals may even become a viable power source for low-power applications, as energy harvesting technologies continue to advance.

\vspace{-.1in}
\subsection{OAM Communication}
Exploiting polarization diversity and OAM mode multiplexing, it has been demonstrated that very high capacity wireless communication systems can be built to work over a distance of a few meters. Several independent data streams transmission over the same spatial wireless channel can be designed which increases the area spectral efficiency by manyfold. The performance is particularly promising at a relatively short distance which can be useful for industrial automation. A mmWave OAM system in \cite{JWang} was reported to have achieved more than 2.5Tbps rate with spectral efficiency of a massive 95.7bps/Hz. This can be a lucrative technology for Industry 4.0 which is envisaged as one of the key 6G use cases.

\vspace{-.1in}
\subsection{Quantum Communications and Networks}
Quantum communication is another promising technology which is likely to contribute considerably towards two essential criteria of 6G, namely extremely high data rate and security \cite{NHoss17}. The inherent security feature of quantum entanglement that cannot be cloned or accessed without tampering it, makes it a rightful technology for 6G and beyond systems. A number of works already demonstrated initial practical implementation of quantum key distribution (QKD) and associated protocols. Another attractive feature of quantum communication is that it is suitable for long distance communication. Nevertheless, current repeater concept is not applicable for quantum communication as entanglement cannot be cloned. Satellites, high altitude platforms and UAVs may be adopted as trusted nodes for key regeneration and redistribution. In terms of designing quantum devices, single photon emitter device has already been realized which currently works at few degrees above the absolute zero temperature. Much work is still needed to make it operate in normal temperatures. It may be a long shot to see much of an impact of quantum communication in 6G.

\section{Conclusions}\label{sec_con}
As 5G is in its final testing phase getting ready for its launch in 2019, discussion has already begun to shape what 6G may be. There are already high-profile initiatives around the world aiming to develop technologies for 6G, such as 6Genesis of Finland and TOWS for 6G LiFi in the UK. While it is too early to define 6G and there are inevitably omissions in any of such discussion, this article has taken a bravery approach to identify possible enabling technologies for 6G and describe the features they bring beyond the capability of 5G. 
Our 6G vision presents a genuine realization of Mitola radio, which has exceptional awareness of the environment (by radar technologies) to make decisions using superb intelligence (by collective AI), with a rich action space to adapt itself in many forms (by intelligent structures, Li-Fi, WPT and energy harvesting, etc.)
Additionally, OAM and quantum communications may appear in 6G if sufficient advances are made. This article has also discussed the limitations of 5G that form the basis of our 6G vision. We have attempted to present the 6G use cases but they are largely natural extensions of the 5G use cases and scenarios. We anticipate that as 5G is being put into service, new use cases will emerge and the industry will find more ambitious and challenging scenarios. Lastly, we like to add that 6G will see a shift from the electronic era of 5G to the optical and photonics era, but this deserves a separate discussion.

\label{sec4}

\ifCLASSOPTIONcaptionsoff
  \newpage
\fi


%
\vspace{-1cm}
\begin{IEEEbiography}{Faisal Tariq} (M'13) received his PhD degree from Open University, UK. He is currently Senior Lecturer in the School of Engineering at University of Glasgow. His main research interests include resource management, 5G/6G system, molecular communication and technologies for assisted living. He is a recipient of best paper award at Wireless Personal Multimedia Conference (WPMC) in 2013. He is currently serving as an editor for Elsevier Journal of Network and Computer Applications.
\end{IEEEbiography}
\vspace{-1cm}
\begin{IEEEbiography} {Muhammad R. A. Khandaker} (S'10-M'13-SM'18)
is currently an Assistant Professor in the School of Engineering and Physical Sciences at Heriot-Watt University. Before joining Heriot-Watt, he worked as a Postdoctoral Research Fellow at University College London, UK, (July 2013 - June 2018). He is an Associate Editor for the IEEE COMMUNICATIONS LETTERS, the IEEE ACCESS and the EURASIP JOURNAL ON WIRELESS COMMUNICATIONS AND NETWORKING.
\end{IEEEbiography}

\vspace{-1cm}
\begin{IEEEbiography}{Kai-Kit Wong} (M'01-SM'08-F'16) is currently a Professor of Wireless Communications with the Department of Electronic and Electrical Engineering, University College London, U.K. He is Area Editor of the IEEE TRANSACTIONS ON WIRELESS COMMUNICATIONS, and Senior Editor of the IEEE COMMUNICATIONS LETTERS and IEEE WIRELESS COMMUNICATIONS LETTERS.
\end{IEEEbiography}
\vspace{-1cm}
\begin{IEEEbiography}{Muhammad Imran}(M'03-SM'12) received the M.Sc. (Distinction) and Ph.D. degrees from Imperial College London, London, U.K., in 2002 and 2007, respectively. He is a Professor of communication systems with the University of Glasgow, UK, and a Vice Dean with Glasgow College UESTC. He is also an Affiliate Professor with the University of Oklahoma, USA, and a Visiting Professor at University of Surrey, UK. He is an Associate Editor for the IEEE COMMUNICATIONS LETTERS, the IEEE ACCESS and the IET Communications Journals.
\end{IEEEbiography}
\vspace{-1cm}

\begin{IEEEbiography}{Mehdi Bennis} (Senior Member, IEEE) is an Associate Professor at the Centre for Wireless Communications, University of Oulu, Finland, and an Academy of Finland Research Fellow. Dr. Bennis has been the recipient of several awards including Fred W. Ellersick Prize and Best Tutorial Prize from the IEEE Communications Society and EURASIP Best Paper Award for the Journal of Wireless Communications and Networks. He is currently an editor of the IEEE TRANSACTIONS ON COMMUNICATIONS.
\end{IEEEbiography}

\vspace{-1cm}
\begin{IEEEbiography}{M\'erouane Debbah} (Fellow, IEEE) received the M.Sc. and Ph.D. degrees from Ecole Nor- male Supérieure Paris-Saclay, France. Since 2007, he has been a Full Professor at CentraleSupelec, Gif-sur-Yvette, France. He was the Director of the Alcatel-Lucent Chair on Flexible Radio. Since 2014, he has been Vice President of the Huawei France R\&D center, Paris, France, and Director of the Mathematical and Algorithmic Sciences Lab. He is a WWRF Fellow and a member of the academic senate of Paris-Saclay. 
\end{IEEEbiography}




\end{document}